\documentstyle[12pt]{article}
\newcommand{\equ}[1]{(\ref{#1})}
\newcommand{\eq}{\begin{equation}}
\newcommand{\eqn}[1]{\label{#1}\end{equation}}
\newcommand{\eea}{\end{eqnarray}}
\newcommand{\eqa}{\begin{eqnarray}}
\newcommand{\eqan}[1]{\label{#1}\end{eqnarray}}
\newcommand{\ba}{\begin{array}}
\newcommand{\ea}{\end{array}}

         \newcommand{\D}{\Delta}

\newcommand{\p}{\psi}

\newcommand{\vf}{{\varphi}}

\newcommand{\journal}[4]{{\em #1~}#2\,(19#3)\,#4;}

\newcommand{\pr}{\journal {Phys. Rev.}}

\newcommand{\prl}{\journal {Phys. Rev. Lett.}}

\newcommand{\cmp}{\journal {Comm. Math. Phys.}}
\newcommand{\cqg}{\journal {Class. Quantum Grav.}}

\newcommand{\np}{\journal {Nucl. Phys.}}

\newcommand{\annp}{\journal {Ann. Phys. (N.Y.)}}

\newcommand{\es}{\\[3mm]}
\makeatletter
\@addtoreset{equation}{section}
\makeatother

\newcommand{\complex}{{\kern .1em {\raise .47ex
\hbox {$\scriptscriptstyle |$}}
    \kern -.4em {\rm C}}}
\newcommand{\real}{{{\rm I} \kern -.19em {\rm R}}}
\newcommand{\rational}{{\kern .1em {\raise .47ex
\hbox{$\scripscriptstyle |$}}
    \kern -.35em {\rm Q}}}
\renewcommand{\natural}{{\vrule height 1.6ex width
.05em depth 0ex \kern -.35em {\rm N}}}
\def\p{\partial}
\newcommand{\reff}[1]{(\ref{#1})}
\newcommand{\fud}[2]  {{\displaystyle{\frac{\delta #1}{\delta #2}}}}
\newcommand{\fudp}[2]  {{\displaystyle{\frac{\p #1}{\p #2}}}}
\newcommand{\tint}{{\displaystyle{\int}} d^3 \! x \, }
\newcommand{\tr}{{\,{\rm {Tr}} \,}}

\newcommand{\sla}{\raise.15ex\hbox{$/$}\kern -.57em}

\newcommand{\twiddle}{\lower.9ex\rlap{$\kern -.1em\scriptstyle\sim$}}
\def\dds#1{
\fud{\Sigma}{#1}
}
\def\ddsc#1{
\fud{\Sigma_c}{#1}
}
\def\ddg#1{
\fud{\Gamma}{#1}
}
\def\dda#1{
\fud{{\Sigma_{\rm inv}}}{#1}
}

\def\ddsigma#1{
\fud{\Sigma}{K_#1}\fud{\Sigma}{#1}
}

\def\ddelta#1#2{
{\cal D}#1\fud{#2}{#1}
}
\def\ee{(\bar\epsilon\epsilon)}
\def\ege{(\bar\epsilon \gamma^\mu \epsilon)}

\def\bsig#1{B_{\Sigma}^{(#1)}}

\def\bL{\bar\Lambda}
\def\bl{\bar\lambda}
\def\bP{\bar\Psi}
\def\bp{\bar\psi}

\def\p{\partial}
\begin{document}
\def\ftoday{{\sl  \number\day \space\ifcase\month
\or Janvier\or F\'evrier\or Mars\or avril\or Mai
\or Juin\or Juillet\or Ao\^ut\or Septembre\or Octobre
\or Novembre \or D\'ecembre\fi
\space  \number\year}}
\titlepage
{

\begin{center}               \

{\huge Algebraic Renormalization of $N=2$\\[2mm] Supersymmetric
Yang-Mills\\[2mm] Chern-Simons Theory in the\\[2mm]
Wess-Zumino Gauge\footnote{
Supported in part by the Swiss National Science Foundation.}}

\vspace{1cm}

{\Large Nicola Maggiore, Olivier Piguet and Mathieu Ribordy}
\vspace{1cm}

{\it D\'epartement de Physique Th\'eorique --
     Universit\'e de Gen\`eve\\24, quai E. Ansermet -- CH-1211 Gen\`eve
     4\\Switzerland}

\end{center}

\vspace{10mm}

\begin{center}
REVISED
\end{center}

\vspace{15mm}

\begin{center}
\bf ABSTRACT
\end{center}

{\it
We consider   a $N=2$ supersymmetric
Yang-Mills-Chern-Simons model, coupled to matter,
in the Wess-Zumino gauge.
The theory is characterized by   a superalgebra which
displays two kinds of obstructions to the closure on the translations:
field dependent gauge transformations,   which give rise to an
infinite algebra, and equations of motion.
The   aim is to put the formalism in a closed form,
off-shell, without introducing auxiliary fields.
  In order to perform that, we collect all
the symmetries of the model into a unique nilpotent
  Slavnov-Taylor operator.
Furthermore, we prove the renormalizability of the model through the
analysis of the cohomology arising from the generalized Slavnov-Taylor
operator.
In particular, we show that the model is free of anomaly.
}

\vfill
\noindent
hep-th/9504065\\
UGVA-DPT-1995/04-886 \hfill April 1995
}
\newpage

\section{Introduction}

In this paper, we consider a Yang-Mills-Chern-Simons
model coupled to matter in a $N=2$ supersymmetric
extension~\cite{Lee,Grigoriev2,Ivanov,Gates1,Gates2,Grigoriev1}
of the usual Yang-Mills--Chern-Simons
theory~\cite{ym-cs,martin1,martin2}
Although a superfield version of the theory does
exist~\cite{Ivanov,Gates2,Grigoriev1}, we shall work in components, moreover
without auxiliary fields, and  in
the Wess-Zumino gauge  (we recall that the Wess-Zumino gauge
fixes all the supergauge of freedom
except the ordinary  one for the vector gauge field).

Because of the complications following from this choice,
namely the nonlinearity
of the supersymmetry transformations and the closure of
the superalgebra only modulo field equations and
field dependent gauge transformations~\cite{so,BM},
it is convenient to follow the approach of~\cite{whimag},
suitable to study
more complicated situations,
in which a simple superspace formalism, either
is not available, as for instance in some extended supersymmetry
theories,  or does not bring simplifications like in
case of broken supersymmetry~\cite{prep}.

However,  mastering
the main complication of the Wess-Zumino gauge, which is the
infinite dimensional algebra spanned by
the field dependent gauge
transformations, turns out to be very difficult~\cite{BM}.

On the other hand, a satisfactory regularization procedure
compatible with both BRS invariance and
supersymmetry is lacking.

All of this makes much advisable
the adoption of the algebraic method
of renormalization~\cite{Piguet}, which is indeed
regularization  scheme independent, and powerful enough to
overcome the intrinsic difficulties of the problem at hand.

Fixing the remaining gauge invariance gives rise
to an additional minor problem: it is not possible
to construct a gauge fixing term which
is invariant under both BRS and supersymmetry.

The difficulties we have just mentioned and
which will be described in Section~\ref{the model}, can
simultaneously be solved
by collecting all the symmetries into a unique nilpotent
operator $\cal D$. In this way,
the original algebra reduces to the simple nilpotency relation
characterizing the generalized BRS operator.
This is achieved in Section~\ref{collecting} and translated in
the form of functional identities in
Sections~\ref{formulation} and \ref{constraints}.
Section~\ref{algebraic}
is devoted to the renormalization of the model,
by finding the counterterms and proving that the model is free of anomalies.

\section{The Model}\label{the model}

The model is described by two   supermultiplets of fields:
the Chern-Simons   supermultiplet (CSM),
which belongs to the adjoint representation of a semi-simple gauge group,
and the matter   supermultiplet (MM),
which is in an arbitrary representation of the
gauge group.

Here are given the fields of   both supermultiplets:
\begin{center}
\begin{tabular}{lrrrr}
CSM: & $A_\mu^a$ & $S^a$    & $\lambda^a$ & $\bar\lambda^a$, \es
MM:  & $A_i$     & $A^{*i}$ & $\psi_i$    & $\bar\psi^i$,
\end{tabular}
\end{center}
where $A^a_\mu$ is the gauge field, $S^a$ is a real scalar field,
$\lambda^a$, $\bar\lambda^a$, $\psi_i$ and $\bar\psi^i$
are Dirac spinors, $A_i$ and $A^{*i}$ are complex scalar fields.

The indices $a$,$b$,$c$ run over the adjoint representation
of the gauge group, whereas $i$,$j$ concern the arbitrary
representation in which the matter fields $A_i$, $\psi_i$ live.
The matrix   generators $(T_a)^i_j$
are supposed to be antihermitian.

In the following, we shall adopt a vector space notation.
For $\psi$, $\psi'$ in an arbitrary representation: $(\psi^*,\psi') \equiv
\psi^{*i}{\psi'}_i$.
Moreover, for $\phi$ in the adjoint representation, we define
$(\psi^*,\phi\psi') \equiv \phi^a (\psi^*,T_a \psi') \equiv
\phi^a \psi^{*i} (T_a)_i^j {\psi'}_j$,
where the $T_a$'s are the generators of the representation where
$\psi, \psi'$ live.
Finally, we use also the notation $\phi \equiv \phi^a \tau_a$, for
$\phi = A_\mu, \lambda$ or $S$, the $\tau_a$'s being the generators of the
group in the fundamental representation, with the normalization
$\tr(\tau^a \tau^b) = \delta^{ab}$.

The action is given by
\begin{equation}
{\Sigma}_{\rm inv} = {\Sigma}_{CS} + {\Sigma}_{VM} +
                    {\Sigma}_{SM}, \label{eq:invaction}
\end{equation}
\noindent where
\begin{equation}
{\Sigma}_{CS} = m \tint \Bigl(
-\textstyle{1\over2}\epsilon^{\mu\nu\rho}
                          \tr\left(F_{\mu\nu} A_\rho
- \textstyle{2 \over 3}gA_\mu A_\nu A_\rho\right)
- 2\tr(\bar\lambda \lambda) -2m \tr S^2 + 4ig (A^*,SA)
                                                      \Bigr),
\label{eq:CS}
\end{equation}
\medskip
\begin{equation}
{\Sigma}_{VM} = \tr\tint \left(
- \textstyle{1 \over 4}F_{\mu\nu} F^{\mu\nu}
+ i\bar\lambda \rlap{D}\,/ \lambda + {1\over 2} D_\mu S D^\mu S
+ 2ig(\bar\lambda \lambda)S \right), \label{eq:VM}
\end{equation}
\medskip
and
\begin{equation}
\begin{array}{rcl}
{\Sigma}_{SM} &=& \tint
\Bigl(
2 (D^\mu A^*,D_\mu A) + i(\bar\psi,\rlap{D}\,/ \psi)
+ ig(\bar\psi,S\psi) \es
&+& 2ig(A^*,\bar\lambda \psi) + 2ig(\bar\psi, \lambda A)
+ 2g^2(A^*,TA)^2 + 2 g^2(SA^*,SA)
\Bigr),
\end{array}  \label{eq:SM}
\end{equation}
where $m$ is a dimensionful coupling constant and   $g$ is
the gauge coupling constant.   We have defined the conjugate spinors as
usual by $\bar\psi = \psi^\dagger\gamma^0$.
The $\gamma$-matrices satisfy a
Clifford algebra and   may be expressed in terms of
the Pauli matrices as
\begin{equation}
\gamma^0  =  \sigma^3, \quad
\gamma^1  =  i\sigma^2, \quad
\gamma^2  =  i\sigma^1.
\end{equation}

The field strength and the covariant derivatives are defined as
\begin{equation}
F_{\mu\nu} = \p_\mu A_\nu - \p_\nu A_\mu +  g[A_\mu, A_\nu],
\end{equation}
\begin{equation}
\begin{array}{rcl}
D_\mu \phi &=& \partial_\mu \phi + g[A_\mu,\phi], \\[1mm]
&&\hbox{for }\phi \hbox{ in the adjoint representation, and} \es
D_\mu \psi &=& \partial_\mu \psi + gA^a_\mu T_a \psi, \\[1mm]
&&\hbox{for }\psi \hbox{ in   the matter field
representation.}
\end{array}
\end{equation}

Besides being gauge invariant, the action~\reff{eq:invaction} is left unchanged
by the $N=2$  supersymmetric transformations
\begin{equation}
\begin{array}{rcl}
\delta A_\mu &=& \bar\epsilon \gamma_\mu \lambda
+ \bar\lambda \gamma_\mu \epsilon \es
\delta\lambda &=& {1 \over 2} F_{\mu\nu}
\gamma^{\mu\nu} \epsilon - iD_\mu S \gamma^\mu \epsilon
-2m S \epsilon + 2ig( A^*,TA) \epsilon \es
\delta S &=& \bar\epsilon\lambda + \bar\lambda\epsilon \es
\delta A &=& \bar \epsilon \psi \es
\delta \psi &=& -2iD_\mu A \gamma^\mu \epsilon +
2ig S A \epsilon
\end{array} \label{susy}
\end{equation}
\noindent where $\epsilon$ is an infinitesimal Dirac spinor parameter.

\noindent Notice that the  infinitesimal supersymmetric variations of the
spinor fields exhibit the nonlinearities arising from the adoption of
the Wess-Zumino gauge.

The commutators of two subsequent infinitesimal supersymmetric variations
of the fields are given by

\begin{equation}
\begin{array}{rcl}
[\delta_1,\delta_2]\Phi &=& 2i \partial_\mu \Phi
 (\bar\epsilon_1\gamma^\mu \epsilon_2-
\bar\epsilon_2 \gamma^\mu \epsilon_1) \es
                           &+& 2i \delta_g^{(\omega)} \Phi \es
                           &+& \hbox{equations of motion,}
\end{array}
\label{comm}
\end{equation}
where $\Phi$ stands for all the fields and $\omega$ is a field
dependent parameter:
\begin{equation}
\omega \equiv A_\mu \ege - S \ee. \label{eq:omega}
\end{equation}

\section{Collecting Symmetries}\label{collecting}

We have seen in the previous section that
the supersymmetry~\reff{susy} was realized nonlinearly
and, consequently, that the commutators did
not close on the translations.
The usual approach when dealing with
algebraic structures like~\reff{comm} is to
add  auxiliary fields in order to put the formalism
off-shell, with possibly the appearance of a central charge
in the algebra of the matter multiplet~\cite{so}.

As well explained in~\cite{BM}, the algebra~\reff{comm}
is infinite dimensional,
because of the field dependent gauge transformations in the result
of the commutators. To control it, one should introduce an
infinite number of external sources with increasing negative dimensions.
This makes the renormalization   problematic.
Our aim, in view of the quantum
extension of the theory, is to construct   an operator $\cal D$
which contains   all the relevant symmetries of the model.
  The algebra will be characterized in a closed way by
demanding the nilpotency of the operator $\cal D$.
With the help of this
operator, it becomes very easy to   calculate
the counterterms as well as the possible anomalies.
In order to construct such a nilpotent operator,
all the symmetries are summed up: in addition to
the supersymmetry and to the BRS symmetry, we must take into
account also the translation invariance of the model. This method~\cite{whimag}
allows
  one to put the formalism off-shell without the help
of   auxiliary fields, whose role
is played, in this formalism, by the external sources coupled to the nonlinear
variations of the quantum fields.


In view of fixing the gauge, we introduce a ghost $c$,
an antighost $\bar c$ and
a Lagrange   multiplier field $b$ implementing the gauge fixing condition
\begin{equation}
\p_\mu A^\mu = 0.
\end{equation}

Once the gauge is fixed, the gauge invariance evolves into the BRS invariance
described by the following action on the quantum   fields:
\begin{equation}
\begin{array}{l}
s A_\mu   =  -(D_\mu c),\quad
s \lambda   =  g[c,\lambda] ,\quad
s S   =  g[c, S], \es
s A   =  gc  A,  \quad
s \psi   =  gc \psi,\es
s c   =  gc^2 ,\quad
s \bar c   =  b,\quad
s b  =  0.
\end{array}  \label{eq:BRSGH}
\end{equation}
  The $s$-operator is nilpotent
\begin{equation}
s^2 =0.
\end{equation}
  We complete   the BRS symmetry with the supersymmetry \reff{susy}
and the translations, collecting these in
a unique operator ${\cal D}$
\begin{equation}
{\cal D} = s + \delta + \xi^\mu \partial_\mu - 2i \ege
                 {\partial \over \partial \xi^\mu},
\label{eq:grandbrs}\end{equation}
  where the infinitesimal parameters $\epsilon$ and $\xi$ (for
supersymmetry and translations, respectively) are promoted to the status
of global ghosts, the spinor $\epsilon$ being now commuting and the
vector $\xi$ anticommuting. The last term in \reff{eq:grandbrs} then ensures
the nilpotency of $\cal D$ -- valid on-shell, i.e. modulo
the equations of motion:
\begin{equation}
{\cal D}^2 = \hbox{equations of motion}.
\end{equation}
The action of   $\cal D$ is explicitly given by
\begin{equation}
\begin{array}{rcl}
{\cal D} A_\mu &=& g[c,A_\mu] - \partial_\mu c
+ \bar\epsilon \gamma_\mu \lambda + \bar\lambda \gamma_\mu \epsilon
+ \xi \p A_\mu \es
{\cal D} \lambda &=& g[c,\lambda]
+ {1 \over 2} F_{\mu\nu} \gamma^{\mu\nu} \epsilon
- iD_\mu S \gamma^\mu \epsilon
-2m g^2 S \epsilon + 2i g(A^*, TA) \epsilon
+ \xi \p \lambda  \es
{\cal D} S &=& g[c,S] + \bar\epsilon\lambda
+ \bar\lambda\epsilon + \xi \p S \es
{\cal D} A &=& gc A + \bar \epsilon \psi  + \xi \p A \es
{\cal D} \psi &=& gc\psi -2iD_\mu A \gamma^\mu \epsilon +
2ig S A \epsilon + \xi \partial \psi \es
{\cal D} \epsilon &=& 0 \es
{\cal D} \xi_\mu &=& -2i(\bar\epsilon \gamma_\mu \epsilon) \es
{\cal D} c &=& gc^2 - 2i\omega + \xi\partial c \es
{\cal D} \bar c &=& b + \xi \p \bar c \es
{\cal D} b &=& 2i(\bar\epsilon\gamma\epsilon)\partial\bar c + \xi\p b,
\end{array}\label{DGH}\end{equation}
  The quantity $\omega$ in the transformation law for $c$
has been given by \reff{eq:omega}.
  The operator $\cal D$ thus defined turns out to be nilpotent on
all fields except the spinors, for which one gets
\begin{equation}
\begin{array}{rcl}
{\cal D}^2 \lambda &=&
\epsilon \left(\left({\dda \lambda}\epsilon\right)
- \left(\bar\epsilon {\dda {\bar\lambda}}\right)\right), \\[4mm]
{\cal D}^2 \psi &=& {\dda {\bp}}(\bar\epsilon \epsilon)
- \gamma_\mu {\dda {\bp}}(\bar\epsilon \gamma^\mu \epsilon).
\end{array} \label{dspinor}
\end{equation}
\noindent   The action~\reff{eq:invaction} is ${\cal D}$-invariant:
\begin{equation}
{\cal D}{\Sigma}_{\rm inv} = 0.
\end{equation}
The canonical dimensions and the   Faddeev-Popov
charges of the quantum fields   and of the global ghosts are listed  in
Table~\ref{table-dim}.
\begin{table}[hbt]
\centering
\begin{tabular}{|c||c|c|c|c|c|c|c|c|c|c|}
\hline
&$A_\mu$&$\lambda$&S&A&$\psi$&$c$&$\bar c$&$b$ &$\epsilon$&$\xi$\\
\hline\hline
$d$&1/2&1&1/2&1/2&1&-1/2&3/2&3/2&-1/2&-1 \\
\hline
$\Phi\Pi$&0&0&0&0&0&1&1&1&1&1 \\
\hline
\end{tabular}
\caption[t1]{Dimensions   $d$ and ghost charges $\Phi\Pi$.}
\label{table-dim}
\end{table}


  In conclusion, the dependence on the gauge transformations
is now removed from the algebra~\reff{comm}.
  This results from the $\omega$-dependent term in the $\cal
D$-transformation of the ghost field $c$, as given in \reff{DGH}.
  Moreover, as anticipated in the Introduction, this goes
not alone: at the same time, we have completed the construction of an on-shell
nilpotent operator ${\cal D}$.

  Finally, we increase the action by a ${\cal D}$-invariant
gauge fixing term
\begin{equation}
\begin{array}{rcl}
{\Sigma}_{gf} &=& {\cal D} \tr \tint \bar c \partial A \es
&=& \tr \tint \left( b \p^\mu A_\mu + \p^\mu \bar c g[c,A_\mu]
+ \bar c \partial^2 c + \partial_\mu \bar c
(\bar\epsilon \gamma^\mu \lambda + \bar\lambda \gamma^\mu \epsilon)
\right),
\end{array}
\end{equation}
  thus escaping from the impossibility of writing a supersymmetry
invariant gauge fixing term.

  Hence the total gauge-fixed action
\begin{equation}
\Sigma_0 = {\Sigma}_{\rm inv} + {\Sigma}_{gf}
\label{sigma0}\end{equation}
is   ${\cal D}$-invariant: ${\cal D}$ is a symmetry of the action.

\section{Slavnov-Taylor Identity and Off--Shell\hfill\break
Formulation}\label{formulation}

The problem of putting the formalism off-shell, namely of getting rid of the
equations of motion in~\reff{dspinor},
is tightly related to the task of writing
the Slavnov-Taylor identity associated to the ${\cal D}$-symmetry~\reff{DGH}.
In order to do that, it is necessary to couple external sources to the
nonlinear ${\cal D}$-variations of the quantum fields.
  But this is not sufficient for putting the formalism
off-shell.
To obtain that, it is necessary to add to the action   also
a term quadratic in the external sources .
\begin{table}[hbt]
\centering
\begin{tabular}{|c||c|c|c|c|c|c|}
\hline
Fields $\varphi$:&$A_\mu$&$\lambda$&$S$&$A$&$\psi$&$c$ \\
\hline
Sources $K_\varphi$:&$\Omega_\mu$&$\bL$&$M$&$U^*$&$\bP$&$L$ \\
\hline
\end{tabular}
\caption[t1]{Notations for the sources.}
\label{table-ext-f}
\end{table}
The total action
\begin{equation}
\Sigma = \Sigma_0 + \Sigma_1 + \Sigma_2,
\end{equation}
  with $\Sigma_0$ defined in \reff{sigma0}, and with
(  See Table~\ref{table-ext-f} for the notations of the
sources)
\begin{equation}
\Sigma_1 = \sum_\varphi  \tint K_\varphi {\cal D} \varphi,
\end{equation}
\begin{equation}
\Sigma_2 = \tr\tint \bigl((\bP \bar\Psi)(\bar\epsilon\epsilon)
-   (\bP \gamma_\mu \bar\Psi)(\bar\epsilon \gamma^\mu \epsilon)
-(\bar\epsilon \Lambda)(\bar\epsilon \Lambda) -(\bL \epsilon)(\bL \epsilon)
- (\bL\epsilon)(\bar\epsilon \Lambda)\bigr), \label{quadr}
\end{equation}
satisfies the Slavnov-Taylor identity
\begin{equation}
{\cal S}(\Sigma) = \tint
\left(
\sum_\varphi
\ddsigma\varphi
+ \tr\left({\cal D}\bar c {\delta \Sigma \over \delta \bar c}
+ {\cal D}b {\delta \Sigma \over \delta b}\right)
\right)
+ {\cal D}\xi {\partial \Sigma \over \partial \xi} = 0.    \label{slavid}
\end{equation}

The linearized Slavnov-Taylor operator given by
\begin{equation}
{\cal B}_\Sigma = \tint
\left(
\sum_\varphi \left({\delta \Sigma \over \delta K_\varphi}
{\delta \over \delta \varphi}
+             {\delta \Sigma \over \delta \varphi}
 {\delta \over \delta K_\varphi}\right)
+ \tr\left(\ddelta{\bar c}{} + \ddelta{b}{} \right)
\right)
+ {\cal D}\xi \fudp{}{\xi}   \label{linsld}
\end{equation}
is  off-shell nilpotent
\begin{equation}
{\cal B}_\Sigma {\cal B}_\Sigma = 0.
\end{equation}

${\cal B}_\Sigma$ acts on the nonspinorial quantum fields in the same way
as ${\cal D}$,
whereas it acts on the spinor fields as follows
\begin{eqnarray}
&&\begin{array}{rcl}
{\cal B}_\Sigma \lambda &=& \dds\bL =
{\cal D}\lambda - 2\epsilon(\bL \epsilon)
- \epsilon(\bar\epsilon \Lambda), \es
{\cal B}_\Sigma \psi &=&  \dds\bP  =
{\cal D}\psi + \Psi(\bar\epsilon\epsilon) -
\gamma_\mu \Psi (\bar\epsilon \gamma^\mu \epsilon).
\end{array} \label{bspinor}
\end{eqnarray}

The addition of the bilinear terms in the sources has the effect of modifying
the transformation laws of the spinor fields in order to get an off-shell
nilpotent operator.  In this sense, the external sources play the same role
as the auxiliary fields.

\section{Constraint Equations}\label{constraints}

In addition to the Slavnov-Taylor identity~\reff{slavid}, the model is
defined
by the following constraints:

1) the $\xi$-equation
\begin{equation}
{\p \Sigma \over \p \xi^\mu} = \Delta_\mu, \label{eq:xi}
\end{equation}
where
\begin{equation}
\begin{array}{rcl}
\Delta_\mu &=& -  \tint \Bigl( \tr\left(
\Omega^\nu \partial_\mu A_\nu + \bar\Lambda \partial_\mu \lambda
+ \partial_\mu \bar\lambda\Lambda + M \partial_\mu S
                   - L \partial_\mu c  \right)\es
 &+& U \partial_\mu A^* + U^* \partial_\mu A
+ \bar\Psi \partial_\mu \psi + \partial_\mu \bar\psi\Psi
\Bigr);
\end{array}
\end{equation}

2) the gauge fixing equation
\begin{equation}
\dds{b} = \partial^\mu A_\mu; \label{eq:b}
\end{equation}

3) the antighost equation
\begin{equation}
\bar{\cal F} \Sigma \equiv
\dds{\bar c} + \partial_\mu \dds{\Omega_\mu}
- \xi \partial \dds{b} = 0, \label{eq:antighost}
\end{equation}
which arises from the commutator of the
linearized Slavnov-Taylor operator ${\cal B}_\Sigma$
with the gauge fixing equation~\reff{eq:b};

4) the ghost equation
\begin{equation}
{\cal F}\Sigma  \equiv  \tint \left( \fud{\Sigma}{c}
          + g[{\bar c},\fud{\Sigma}{b}] \right)
        = g\Delta, \label{ghost}
\end{equation}
where
\begin{equation}
\begin{array}{rcl}
\Delta &=& \tint \Bigl(
[\eta, A] + [M,S] + [\bL,\lambda] + [\bl,\Lambda] \es
 &+& AU^* + UA^* + \bP\psi + \bp\Psi + [L, c] \Bigr) .
\end{array}
\end{equation}

Notice that all the above constraints but the
antighost equation~\reff{eq:antighost},
have the form of symmetries broken by terms which are linear in the quantum
fields, therefore not receiving radiative corrections.

\section{Algebraic Renormalization}\label{algebraic}

Calculations up to two loops~\cite{Gates2,Grigoriev1}
  have  shown the vanishing of the $\beta$-functions
associated to the coupling constants $g^2$ and $m$. From these results, a
finiteness conjecture has been inferred, which is still far from having
being transformed into a proof to all orders. The formulation presented
in the previous sections is the only one suitable for a discussion of the
renormalization to all
orders performed in the Wess-Zumino gauge~\cite{whimag}.
The reason for that is twofold.
First of all, the lack of a coherent
regularization scheme entails the algebraic approach.
Secondly, the nonlinearities of the   supersymmetry
transformations~\reff{susy},
together with the consequent non-closure of the algebra, render problematic
a separate analysis of the BRS symmetry and   of the supersymmetry.

  The algebraic study of the renormalizability of the model develops into
two steps.

First, we will find the counterterms through the study of the stability of
the classical action and we will show
that the radiative corrections can be reabsorbed by a
redefinition of the fields and of the coupling constants of the theory.

Next, we will compute the possible anomaly
through a cohomological analysis of the
nilpotent operator ${\cal B}_\Sigma$.
We will use the filtration method
developed by Dixon~\cite{Dixon}, which will be explained in more details later.
It basically consists in making a judicious choice of a filtration operator,
which  will lead to a simplification of the cohomology problem.

\subsection{  Counterterms}

In order to study the stability of the model under radiative
  corrections,
we perturb the classical action $\Sigma$ by a functional $\Sigma_c$
\begin{equation}
\Sigma \longrightarrow \Sigma + \eta \Sigma_c,
\end{equation}
where $\eta$ is an infinitesimal parameter. We then ask that the perturbed
action satisfies all the symmetries and constraints defining the theory,
which, at first order in $\eta$, imply the following conditions on $\Sigma_c$:

1) from the $\xi$-equation:
\begin{equation}
{\p \Sigma_c \over \p \xi^\mu} = 0      \label{xic}
\end{equation}

2) from the gauge condition:
\begin{equation}
\ddsc{b} = 0                        \label{gcc}
\end{equation}

3) from the antighost equation:
\begin{equation}
\ddsc{\bar c} + \partial_\mu \ddsc{\Omega_\mu} = 0   \label{aghc}
\end{equation}

4) from the ghost equation:
\begin{equation}
\tint \ddsc{c} =0                                    \label{ghc}
\end{equation}

5) from the Slavnov-Taylor identity:
\begin{equation}
{\cal B}_\Sigma \Sigma_c = 0                        \label{slavc}
\end{equation}

The conditions~\reff{xic} and~\reff{gcc} mean that the counterterm $\Sigma_c$
does depend neither on the global ghost $\xi^\mu$ nor on the Lagrange
multiplier $b$. Moreover, the constraints~\reff{aghc} and~\reff{ghc}
are satisfied by a functional depending on $\bar c$ and $\Omega^\mu$
only through the combination
\begin{equation}
\eta^\mu \equiv \p \bar c  +  \Omega^\mu         \label{etaeq}
\end{equation}
and depending on the ghost $c$ only if differentiated
\begin{equation}
c_\mu \equiv \p_\mu c.
\end{equation}
The last condition~\reff{slavc},
which derives from imposing the Slavnov-Taylor
identity on the perturbed action, due to the nilpotency of the linearized
Slavnov-Taylor operator ${\cal B}_\Sigma$,
constitutes a cohomology problem whose
solution must be found within   the space of functionals having
canonical dimensions up to three, vanishing Faddeev-Popov charge and
satisfying all   the previous constraints. Therefore,
the most general solution is
\begin{equation}
\Sigma_c = \hat\Sigma_c + {\cal B}_\Sigma \tilde \Sigma_c.
\end{equation}
The functional $\hat\Sigma_c$ is a nontrivial BRS cocycle, in the
sense that it cannot be written as a ${\cal B}_\Sigma$-variation.
It represents the cohomology of the
${\cal B}_\Sigma$ operator in the sector of   zero $\Phi\Pi$ charge,
and it corresponds to renormalizations of the physical parameters of
the theory.
The trivial cocycle ${\cal B}_\Sigma \tilde \Sigma_c$, on
the other hand, stands for   (unphysical) field amplitude renormalizations.
The results stated in the Appendix insure that the counterterm is
at least of order~$g^2$. This leads to the following expression for the
most general counterterm:
\begin{equation}
\Sigma_c = Z_m S_{CS},
\label{contre}
\end{equation}
where $Z_m$
is an arbitrary constant proportional to  $g^2$ and
$S_{CS}$ is
given by~\reff{eq:CS}.
We see from~\reff{contre}, that {\it a priori}
only the dimensionful constant~$m$ can get radiative corrections.
This means that the beta
function related to the gauge coupling constant~$g^2$
is vanishing to all orders of perturbation theory, and the anomalous
dimensions of the fields as well. We stress that this is a purely algebraic
result, valid to all orders of perturbation theory,
to be compared with the results given in~\cite{Grigoriev1},
which led  to the finiteness conjecture corresponding
to a theory whose   counterterm is a trivial BRS cocycle.
To our knowledge, up to now, no algebraic proof has been given of this
property. Here we   can just state that the radiative
corrections can be reabsorbed through a redefinition of the topological
mass only, thus   concluding the first part of the
renormalizability proof.

\subsection{Anomaly}

To complete the proof of the renormalizability of the model, we have to
show that all the symmetries defining the theory can be extended to the
quantum level, or, in other words,  we must prove that it is possible
to define a quantum vertex functional
\begin{equation}
\Gamma = \Sigma + O(\hbar)
\end{equation}
such that
\begin{equation}
{\p \Gamma \over \p \xi^\mu} = \Delta_\mu, \quad
\ddg{b} = \partial^\mu A_\mu, \quad
\bar{\cal F}\Gamma  = 0, \quad
{\cal F}\Gamma = g\Delta, \quad
\label{qqq}\end{equation}
\begin{equation}
{\cal P}_\mu \Gamma = {\cal W}_{rig} \Gamma = 0   \label{trans}
\end{equation}
\begin{equation}
{\cal S}(\Gamma) = 0, \label{slavqu}
\end{equation}
where   $\bar{\cal F}$, $\cal F$ are defined by
\reff{eq:antighost},
\reff{ghost}, and
${\cal P}_\mu$ and ${\cal W}_{rig}$ are the Ward operators for
translations and rigid gauge transformations respectively.

On the one hand,
the extension of the $\xi$-equation~\reff{eq:xi},
the gauge condition~\reff{eq:b},
the antighost equation~\reff{eq:antighost} and the  ghost equation~\reff{ghost}
to their quantum counterparts is trivial and we refer to~\cite{Piguet} for the
details of the proof. The quantum implementation~\reff{slavqu} of the
classical Slavnov-Taylor identity~\reff{slavid},
on  the other hand, requires some care.
The rest of this section will be devoted to this end.

According to the quantum action principle~\cite{Piguet,qap},
the Slavnov-Taylor identity gets a
quantum breaking
\begin{equation}
{\cal S}(\Gamma) = \Delta \cdot \Gamma
\label{anomal-slavnov}\end{equation}
which, at lowest order in $\hbar$, is a local integrated functional with
canonical dimension three and Faddeev-Popov charge one
\begin{equation}
\Delta \cdot \Gamma  = \Delta + O(\hbar\Delta).
\end{equation}

The fact that $\Gamma$ satisfies the identities~\reff{qqq} to~\reff{trans} and
the algebraic relations  valid for any functional $\gamma$:
\begin{equation}
\fud{}{b}{\cal S}(\gamma)
- {\cal B}_\gamma \left(\fud{\gamma}{b} - \p A\right) =
       \bar{\cal F} \gamma
\end{equation}
\begin{equation}
{\p \over \p \xi^\mu} {\cal S}(\gamma)
+ {\cal B}_\gamma \left({\p \gamma \over \p \xi^\mu}
- \Delta_\mu\right) = {\cal P}_\mu \gamma
\end{equation}
\begin{equation}
\bar{\cal F}{\cal S}(\gamma) +  {\cal B}_\gamma \bar{\cal F} \gamma = 0
\end{equation}
\begin{equation}
{\cal F}{\cal S}(\gamma)
+ {\cal B}_\gamma({\cal F}\gamma - \Delta) = {\cal W}_{rig}\gamma
\end{equation}
\begin{equation}
{\cal B}_\gamma {\cal S}(\gamma)  = 0
\end{equation}
imply the following consistency conditions on the breaking $\Delta$:
\begin{equation}
\fud{\Delta}{b} =  0,\quad
{\p \Delta \over \p \xi^\mu} = 0,\quad
\bar{\cal F} \Delta = 0,\quad
{\cal F} \Delta = 0                        \label{fant}
\end{equation}
\begin{equation}
{\cal B}_\Sigma \Delta = 0                \label{brsl}
\end{equation}
Notice that the consistency
conditions~\reff{fant} and~\reff{brsl} formally coincide
with the relations determining the counterterm. The difference is that now
the solution must belong to the space of functionals having Faddeev-Popov
charge
one instead of zero. Therefore, the first four conditions tell us that the
breaking $\Delta$ does depend neither on $\xi^\mu$ nor on $b$, that $\bar c$
and $\Omega^\mu$ appear  only in the combination $\eta^\mu$~\reff{etaeq} and
that the ghost  must always be differentiated.

The last consistency condition~\reff{brsl}
is often called the Wess-Zumino consistency
condition.   Like  the corresponding one
in the zero-$\Phi\Pi$ sector~\reff{slavc},
  solving it is  a cohomology problem.

We will show that the cohomology of the ${\cal B}_\Sigma$ operator in the
$\Phi\Pi$ charge one sector is empty, namely that the more general solution
of~\reff{brsl} is
\begin{equation}
\Delta = {\cal B}_\Sigma \hat \Delta, \label{cohomeq}
\end{equation}
  with $\hat\Delta$ obeying the constraints \reff{xic} to
\reff{ghc}.   This will entail the possibility of
absorbing the breaking $\Delta$ as a counterterm
$-\hat\Delta$,
 leading thus to the desired conclusion
concerning the absence of anomalies.

To analyze the cohomology of the
${\cal B}_\Sigma$ operator~\reff{linsld}, we adopt
the strategy of~\cite{BBBCD} and~\cite{Dixon}, which consists into
passing from functionals to functions. This corresponds in practice into
translating the functional operator ${\cal B}_\Sigma$, which acts on the
space of local functionals
satisfying~\reff{fant}, into an ordinary differential
operator $B_\Sigma$, which acts on a space of functions $\Delta(x)$ with
dimension three, $\Phi\Pi$ charge one and also restricted to be invariant
under~\reff{brsl}.
Consequently, the cohomology problem~\reff{brsl} can be
cast into the following local identity:
\begin{equation}
B_\Sigma \Delta(x) + d\Delta'(x) =0, \label{simo}
\end{equation}
where $d$ is the exterior derivative: $d^2=0$, and $\Delta(x)$ is
a 3-form defined by $\Delta=\int \Delta(x)$.

Given the functional operator ${\cal B}_\Sigma$, the form of the corresponding
differential operator $B_\Sigma$ is straightforward, provided we consider
as independent the fields and their derivatives.
The results of~\cite{Dixon} insure that
the cohomology of $B_\Sigma$
is isomorphic to a subspace of the cohomology of $\bsig0$.
$\bsig0$ is obtained from
$B_\Sigma$ by making an arbitrary filtration on the fields by means of an
operator ${\cal N}$, and taking
the lowest order
\begin{equation}
B_\Sigma = \sum_{n=0}^N \bsig{n}.
\end{equation}
Let us analyze the identity~\reff{simo}, written for~$\bsig0$:
\begin{equation}
\bsig0\Delta^1_3 + d\Delta^2_2 =0, \label{lad1}
\end{equation}
where, as usual, $\Delta^q_p$ denotes a $p$-form with ghost number~$q$.
Due to the anticommutation relation $[\bsig0, d]=0$ and to the vanishing
of the cohomology of~$d$~\cite{Piguet}, the equation~\reff{lad1} gives
rise to a ladder of descent equations
\begin{equation}
\bsig0\Delta^2_2 + d\Delta^3_1 = 0 \label{lad2}
\end{equation}
\begin{equation}
\bsig0\Delta^3_1 + d\Delta^4_0 = 0 \label{lad3}
\end{equation}
\begin{equation}
\bsig0\Delta^4_0 = 0. \label{lad4}
\end{equation}

It is evident that a good choice of the filtration leads to
an operator $\bsig0$ whose cohomology space is easy to find.
The filtration we adopt
is associated to the operator ${\cal N}$ which
assigns the weights displayed\footnote{Due to the power counting
restrictions, only the (differentiated)
fields up to a certain dimension are considered.}
 in Table~\ref{weights},
where $S_{\mu\nu}\equiv\partial_\mu A_\nu +\partial_\nu A_\mu$,
$G_{\mu\nu}\equiv\partial_\mu A_\nu -\partial_\nu A_\mu$,
$\eta_{\mu\nu} \equiv  \epsilon_{\mu\nu\rho} \eta^\rho$,
$\eta_{\mu\nu\rho}\equiv \p_\mu \eta_{\nu\rho}
+ \p_\nu \eta_{\rho\mu} + \p_\rho \eta_{\mu\nu}$,
and $L_{\mu\nu\rho} \equiv \epsilon_{\mu\nu\rho} L$.
\begin{table}[hbt]
\centering
\begin{tabular}{|c|c|c|c|c|c|c|c|c|c|c|}
\hline
$\epsilon$ & $A_\mu$ & $S_{\mu\nu}$ & $G_{\mu\nu}$ & $\partial_\mu
S_{\nu\rho}$
& $\partial_\mu G_{\nu\rho}$ & $S$ & $\partial_\mu S$ &
$\partial_\mu\partial_\nu S$ & $\lambda$ & $\partial_\mu\lambda$
\\ \hline
2&2&2&1&2&2&1&2&2&1&2
\\ \hline\hline
$\p_\mu\p_\nu\lambda$& $A$ & $\partial_\mu A$ & $\partial_\mu\partial_\nu A$ &
$\psi$ & $\partial_\mu\psi$ & $\p_\mu\p_\nu\psi$ &
$\eta_{\mu\nu}$ & $\eta_{\mu\nu\rho}$ & $\Lambda$
& $\partial_\mu\Lambda$
\\ \hline
3&3&3&1&1&1&1&1&1&1&2
\\ \hline\hline
$M$ & $\partial_\mu M$ & $U$ & $\partial_\mu U$ & $\Psi$ & $ \partial_\mu\Psi$
&
$L_{\mu\nu\rho}$ & $c$ & $c_\mu$ & $\partial_\mu c_\nu$ &
$\partial_\mu\partial_\nu c_\rho$
\\ \hline
1&2&1&1&1&1&1&1&2&2&2
\\ \hline
\end{tabular}
\caption[t3]{Weights.}
\label{weights}
\end{table}

The nonvanishing $B_\Sigma^{(0)}$-transformations of the fields
are
\begin{equation}
\begin{array}{rclrclrcl}
\bsig0 A_\mu &=& - c_\mu, &
\bsig0 S_{\mu\nu} &=& - 2 \p_\mu c_\nu, &
\bsig0 \p_\mu S_{\nu\rho} &=&
- 2 \p_\mu \p_\nu c_\rho,
\\[2mm]
\bsig0 A &=& \bar\epsilon\psi, &
\bsig0 \partial_\mu A &=& \bar\epsilon\partial_\mu\psi, &
\bsig0 \eta_{\mu\nu} &=& -m G_{\mu\nu},
\\[2mm]
\bsig0 M &=& -4m^2 S, &
\bsig0 \p_\mu M &=& -4m^2 \p_\mu S, &
\bsig0 \Lambda &=& - 2m \lambda,
\\[2mm]
\bsig0 \p_\mu \Lambda &=& - 2m \p_\mu \lambda, &
\bsig0 U &=& -2 \p^2 A, &
\bsig0 \p_\mu U &=& -2 \p_\mu \p^2 A,
\\[2mm]
\bsig0 \Psi &=& i\gamma^\mu \p_\mu \psi, &
\bsig0 \p_\mu \Psi &=& i \gamma^\nu \p_\nu \p_\mu \psi, &
\bsig0 L_{\mu\nu\rho} &=& - \eta_{\mu\nu\rho}.
\end{array}
\end{equation}

As a general property, $\bsig0$ is nilpotent.
As a consequence of the particular
filtration chosen, we see that most of the
fields forming the basis for the space
$\Delta(x)$ transform as BRS doublets ($\bsig0 u=v$, $\bsig0 v=0$), which
therefore do not appear in the local cohomology
of $\bsig0$. We can then restrict the computation of the cohomology to the
space spanned by the polynomials of the fields
$c$, $\p_\mu(\p_\nu A_\rho - \p_\rho A_\nu)$,
$\psi$, $\p_\mu\psi$, $\p_\mu\p_\nu\psi$, $A$, $\p_\mu A$,
$\p_\mu\p_\nu A$, $\epsilon$
and their complex conjugates (the number of derivatives being
limited by the powercounting bound on the dimension of
$\Delta$).

The last of the descent equations~\reff{lad4} is a problem of local
cohomology, for which we can therefore use the result just stated.
The most general scalar with dimension bounded by zero and $\Phi\Pi$
charge four, candidate for the nontrivial solution of~\reff{lad4}, is
\begin{equation}
\begin{array}{lr}
a_1g^2(T^{abcd})^i_jc^ac^bc^cc^dA_iA^{\ast j}
+a_2g^2\ee(T^{ab})^i_jc^ac^bA_iA^{\ast j}&\\[2mm]
+a_3g^2(\bar\epsilon\epsilon)^2A_iA^{\ast i}
+a_4g^4T^{abcd}c^ac^bc^cc^d + c.c.
\end{array}
\end{equation}
where $T^{abcd}$, $(T^{abcd})^i_j$ and $(T^{ab})^i_j$ are invariant
tensors, and $a_i$ are arbitrary coefficients. The invariance under~$\bsig0$
imposes that all the~$a_i$ vanish, but~$a_4$. No invariant tensor of
degree four exists, which is completely antisymmetric in its indices,
therefore the only solution is the trivial one
\begin{equation}
\Delta^4_0 = \bsig0\Delta^3_0.
\label{636}\end{equation}
Substituting the expression for $\Delta^4_0$ in~\reff{lad3} transforms
the problem of local cohomology modulo~$d$ in a pure local one
\begin{equation}
\bsig0(\Delta^3_1-d\Delta^3_0)\equiv\bsig0\hat\Delta^3_1=0.
\label{637}\end{equation}
With the aforementioned fields spanning the cohomology space of $\bsig0$,
it is possible to form the following vector, with dimension bounded by
one and~$\Phi\Pi$ charge three:
\begin{equation}
a_1g^2(T^{ab})^i_jc^ac^bA_i(\bar\psi^j\gamma_\mu\epsilon)
+a_2g^2(\bar\epsilon\gamma_\mu\epsilon)(\bar\epsilon\psi_i)A^{\ast i}
+a_3g^3(T^a)^i_j(\bar\epsilon\gamma_\mu\epsilon)c^aA_iA^{\ast j}+c.c.
\end{equation}
For such term to be invariant, it must be $a_1=a_3=0$, and the term left
is a trivial cocycle. This implies that, again,
the most general solution of~\reff{637} is
\begin{equation}
\hat\Delta^3_1=\bsig0\Delta^2_1.
\label{638}\end{equation}
The equation~\reff{lad2} becomes a problem of local cohomology as well
\begin{equation}
\bsig0\Delta^2_2+d(d\Delta^3_0+\bsig0\Delta^2_1)=
\bsig0(\Delta^2_2-d\Delta^2_1)\equiv\bsig0\hat\Delta^2_2=0.
\label{639}\end{equation}
The most general 2-form with dimension bounded by two and ~$\Phi\Pi$
charge two, candidate for being a nontrivial solution of~\reff{639} is
\begin{equation}
\begin{array}{lr}
a_1g^2(T^{ab})^i_j\epsilon_{\mu\nu\rho}c^ac^bA_i(\p^\rho A^{\ast j})
+a_2g^2\epsilon_{\mu\nu\rho}(\bar\epsilon\gamma^\rho\psi_i)(\bar\psi^i\epsilon)
+a_3g^2\epsilon_{\mu\nu\rho}(\bar\epsilon\epsilon)A_i(\p^\rho A^{\ast i})
&\\[2mm]
+a_4g^2(\bar\epsilon\gamma_\nu\epsilon)A_i(\p_\nu A^{\ast i})
+a_5g^2(T)^{ij}_{kl}\epsilon_{\mu\nu\rho}(\bar\epsilon\gamma^\rho\epsilon)
    A_iA_jA^{\ast k}A^{\ast l}
+a_6g^3(T^a)^i_j\epsilon_{\mu\nu\rho}c^aA_i(\bar\psi^j\gamma^\rho\epsilon)
&\\[2mm]
+a_7g^3\epsilon_{\mu\nu\rho}(\bar\epsilon\gamma^\rho\epsilon)A_iA^{\ast i}
+c.c.
\end{array}
\end{equation}
It is easily seen that only the second term is invariant under~$\bsig0$.
Since it can be written as a trivial cocycle, the
equation~\reff{lad1}
also becomes a problem of local cohomology:
\begin{equation}
\bsig0\Delta^1_3+d(d\Delta^2_1+\bsig0\Delta^1_2)
=\bsig0(\Delta^1_3-d\Delta^1_2)\equiv\bsig0\hat\Delta^1_3=0.
\end{equation}
The scalar candidate for belonging to the local cohomology in the~$\Phi\Pi$
sector of charge one, with dimension bounded by three is
\begin{equation}
\begin{array}{lr}
a_1g^2(\bar\epsilon\gamma^\mu\p_\mu\psi_i)A^{\ast i}
+a_2g^2(\bar\epsilon\gamma^\mu\psi_i)(\p_\mu A^{\ast i})
+a_3g^2(T)^{ik}_{jl}(\bar\epsilon\psi_i)A^{\ast j}A_kA^{\ast l}
&\\[2mm]
+a_4g^3(T^a)^{ij}_{kl}c^aA_iA_jA^{\ast k}A^{\ast l}
+a_5g^4(\bar\epsilon\psi_i)A^{\ast i}
+a_6g^5(T^a)^i_jc^aA_iA^{\ast j} +c.c.
\end{array}
\end{equation}
After imposing the $\bsig0$ invariance, what remains is
$\bsig0 (a_5g^4A_iA^{\ast i})$.
This evidently implies that the local cohomology modulo~$d$
of $\bsig0$ in the sector of $\Phi\Pi$-charge one
 is empty, which  in turn entails the vanishing of the
cohomology of the functional operator ${\cal B}_\Sigma$. We therefore
conclude that it is possible to construct a quantum vertex functional $\Gamma$
satisfying the Slavnov-Taylor identity~\reff{slavqu} to all
orders of perturbation
theory, without the presence of any anomaly.

\section{  Conclusion}\label{conclusion}
We were able to unify into a single nilpotent generalized BRS
operator $\cal D$
the gauge invariance of the $2+1$-dimensional,
$N=2$ supersymmetric
Yang-Mills--Chern-Simons model in the Wess-Zumino gauge,
without auxiliary fields,
together with supersymmetry and translation invariance.
As a main result, this led to a finite algebra, closed off-shell
after the introduction of the external fields associated to the
nonlinear symmetry transformations, and particularly
after the introduction of quadratic terms in these external
fields.

The use of the algebraic
method of renormalization together with the study of the
cohomology of the operator $\cal D$,
has allowed us to show that the model
is perturbatively renormalizable to all
orders. First, anomalies have been proven to be absent.
Next, the
study of the possible counterterms has led to the conclusion
that the theory is multiplicatively renormalizable,
namely that the counterterms can be reabsorbed by a redefinition of
the topological mass only. We were able to prove
in a very simple way, avoiding any Feynman graph computation,
that, to all orders of perturbation theory,
the gauge coupling constant does not get radiative corrections and that
the fields have no anomalous dimensions.
This latter result, on the one hand,
is weaker than the known ultraviolet
finiteness~\cite{Gates2,Grigoriev1} of the
model, i.e. the nonrenormalization of its
coupling constant $and$ mass. On the other hand, our approach
leads to a statement valid to all orders, whereas the finiteness,
at the best of our knowledge, has been completely proven up
to the two loops order only.

Our work extends the renormalizability of the
Yang-Mills--Chern-Simons theory
-- which was in fact shown to be ultraviolet
finite~\cite{martin2} -- to its $N=2$
supersymmetric generalization.
%
%
\vskip5mm
\noindent {\bf Acknowledgments}
Sylvain Wolf is gratefully acknowledged for his constructive observations
and for a critical reading of the manuscript.
\renewcommand{\thesection}{A}
\section*{Appendix: Power Counting}
The degree of divergence of a 1-particle irreducible Feynman
graph $\gamma$ is given by
\eq
d(\gamma) = 3-\sum_\varphi d_\varphi N_\varphi - \frac{1}{2}g.
\eqn{div-degree}
Here $N_\varphi$ is the number of external lines
of $\gamma$ corresponding to the field $\varphi$
(the global ghost $\epsilon$ being considered as a field, too),
$d_\varphi$ is the dimension of $\varphi$ as given in
Table~\ref{table-dim}, and $N_g$ is the power of
the coupling constant $g$ in the integral corresponding to the
diagram $\gamma$.

Let us recall that a nonnegative value of $d(\gamma)$ corresponds to
ultraviolet divergence. The dependence on the coupling constant, i.e. on
the perturbative order, is
characteristic of a superrenormalizable theory. The equivalent expression
\eq
d(\gamma) = 4-\sum_\varphi \left(d_\varphi+\frac{1}{2}\right)
          N_\varphi - L,
\eqn{div-degree-l}
where $L$ is the number of loops of the diagram, shows that only graphs
up to the too-loop order are divergent.

In order to apply the known
results on the quantum action principle~\cite{Piguet,qap}
to the present
situation, one may consider $g$ as an external field of dimension $1/2$.
Including it in the summation under $\vf$, \equ{div-degree} gets the
same form as in a strictly renormalizable theory:
\eq
d(\gamma) = 3-\sum_\varphi d_\varphi N_\varphi.
\eqn{div-degree-g}
Thus, including the dimension of $g$ into the calculation, we may state
that the dimension of
the counterterms of the action is bounded by 3.
But, since they are generated by loop graphs, they are of order 2 in $g$
at least. This means that, not taking now into account
the dimension of $g$, we can conclude that their
real dimension is bounded by 2.

In the same way, we arrive at the result that the Slavnov-Taylor breaking
$\D$ defined in \equ{anomal-slavnov} has a dimension bounded by 3 (counting
the dimension  of $g$), bounded by 2 if we don't count the dimension of
$g$, since, being produced by  the
radiative corrections, it is of order $g^2$ at least, too.

\end{document}